\documentclass[aps,pra,preprint,showpacs,amsmath,nofootinbib,superscriptaddress]{revtex4}

\usepackage{graphicx}
\usepackage{ulem}
\usepackage{amsfonts}
\usepackage{amsmath}
\usepackage{amssymb}
\usepackage{mathrsfs}

\newcommand{\ab}{\alpha\beta}
\newcommand{\partialP}{\frac{\partial}{\partial P}}
\newcommand{\partialR}{\frac{\partial}{\partial R}}
\newcommand{\partialt}{\frac{\partial}{\partial t}}
\newcommand{\half}{\frac{1}{2}}

\newcommand{\aabb}{\alpha\alpha',\beta\beta'}

\newcommand{\coswt}{\cos{(\omega t)}}

\newcommand{\qomega}{\frac{\Omega^2}{4}}

\newcommand{\Vb}{V_{b}}

\newcommand{\gcossq}{g^2\cos^2(\omega t)}

\newcommand{\gcos}{g\coswt}

\newcommand{\iohbar}{\frac{i}{\hbar}}
\newcommand{\alal}{\alpha\alpha'}

\newcommand{\rhowalal}{\rho_{\rm W}^{\alal}}
\newcommand{\tomega}{\tilde\omega}
\newcommand{\rhooo}{\rho_{\rm W}^{11}}

\newcommand{\rhoto}{\rho_{\rm W}^{21}}
\newcommand{\rhott}{\rho_{\rm W}^{22}}
\newcommand{\etaoo}{\eta_{\rm W}^{11}}
\newcommand{\etato}{\eta_{\rm W}^{21}}

\newcommand{\etatt}{\eta_{\rm W}^{22}}
\newcommand{\omegad}{\omega_{d}}
\newcommand{\gcoswt}{g\cos(\omegad t)}
\newcommand{\derivd}{\textrm{d}}

\begin{document}

\title{Quantum dynamics of a plasmonic metamolecule
with a time-dependent driving}

\author{Daniel A. Uken}
\email{danuken@gmail.com}

\affiliation{
School of Chemistry and Physics, University of KwaZulu-Natal in Pietermaritzburg,
Private Bag X01, Scottsville 3209, South Africa}

\author{Alessandro Sergi
}
\email{sergi@ukzn.ac.za}

\affiliation{
School of Chemistry and Physics, University of KwaZulu-Natal in Pietermaritzburg,
Private Bag X01, Scottsville 3209, South Africa}
\affiliation{
KwaZulu-Natal Node,
National Institute for Theoretical Physics (NITheP), South Africa}

\begin{abstract}
We simulate the dynamics of a quantum dot coupled to the
single resonating mode of a metal nano-particle.
Systems like this are known as metamolecules.
In this study, we consider a time-dependent
driving field acting onto the metamolecule.
We use the Heisenberg equations of motion for the entire system,
while representing the resonating mode in Wigner phase space.
A time-dependent basis is adopted for the quantum dot.
We integrate the dynamics of the metamolecule for a range of coupling strengths
between the quantum dot and the driving field, while restricting
the coupling between the quantum dot and the resonant mode
to weak values.
By monitoring the average of the time variation of the energy
of the metamolecule model, as well as the coherence and
the population difference of the quantum dot,
we observe distinct non-linear behavior in the case of strong coupling to
the driving field.
\end{abstract}

\maketitle

\section{Introduction}

Quantum plasmonics
is a relatively new area of research that studies the interaction
of surface plasmons with
quantum emitters~\cite{plasmonics-review-tame,plasmonics-chen,plasmonics-zuloaga,plasmonics-marinica,plasmonics-bouillard}.
The surface plasmon and the quantum emitter constitute what is known
as a metamaterial or metamolecule. Such plasmonic metamaterials,
at variance from those that operate
in the microwave regime~\cite{metamaterials-valentine,metamaterials-pratibha,metamaterials-leonhardt}, require the downscaling
to nanometers.
Metal nano-particles (MNP's) seem to be an ideal candidate in order to build
metamaterials at optical frequencies~\cite{mnp-lance-kelly,mnp-lal,mnp-nehl,mnp-pelton}. They also have the convenient
property that their resonant frequency can be tuned
by changes in their geometries~\cite{mnp-lance-kelly}. 
A typical example of such systems is provided by metamolecules
comprising MNP's coupled to quantum dots (QD's)~\cite{savasta,plasmonics-tame}.
These advances strengthen the need for techniques capable of simulating the
dynamics of MNP's coupled to QD's.
In the past two decades, much progress has been made in the development of methods
for the simulation of quantum dynamics, involving path integral formulations~\cite{path},
mean field approximations~\cite{meanfield}, semiclassical approximations and surface-hopping
schemes~\cite{tully,miller,pechukas1,pechukas2,heller,shenvi}. 
An alternative approach is provided by the partial Wigner representation 
of quantum mechanics~\cite{wigner1,wigner2,wigner3,wigner4,wigner5,wigner6}.
In such a representation, 
exact algorithms for subsystems embedded in harmonic environments
can be developed (when the coupling to the environment is bilinear)~\cite{num-app-quant}.

In this paper, we present a method for simulating,
within the partial Wigner representation of quantum mechanics,
a MNP-QD metamolecule subject to an external driving field.
To this end, we use a piece-wise deterministic algorithm
which utilizes a time-dependent basis.
We study the dynamics of the population difference of the QD,
in the case of weak
coupling to the MNP, when the QD is subjected to an external driving field
of varying strengths. 
By monitoring the average of the time variation of the energy
of the metamolecule model, as well as the coherence and 
the population difference of the quantum dot,
we observe distinct non-linear behavior in the case of strong coupling to
the driving field.
For validating our approach,
we perform a comparison with the results obtained by
employing a brute-force numerical approach, which uses
a discretized phase-space grid and deals with
the partial differential equation system associated
with the evolution of the density matrix in the partial Wigner representation. 

The structure of the paper is as follows. In Sec.~\ref{sec:sec2},
the formalism of quantum dynamics in the partial Wigner representation
is presented.
Section~\ref{sec:sec3} 
outlines the generalization of the formalism to a time-dependent Hamiltonian.
Section~\ref{sec:sec4} introduces the model for the QD coupled to the MNP.
In the same section,
a brief outline of the propagation algorithm is discussed.
In Sec.~\ref{sec:sec5}, the results of the numerical simulations
are presented.
Finally in Sec.~\ref{sec:sec6},
we give our conclusions.
Appendix~\ref{app:app1} details the adimensional coordinates
used in our study while App.~\ref{app:app2} outlines 
the phase-space-grid algorithm used to verify the results.


\section{Quantum dynamics in the partial Wigner representation}
\label{sec:sec2}

Consider a system defined by the following
Hamiltonian operator:

\begin{eqnarray}
\hat{H} = \hat{H}_{\rm S}(\hat{r},\hat{p}) + \hat{H}_{\rm B}(\hat{R},\hat{P}) + 
\hat{H}_{\rm C}(\hat{r},\hat{R}) +
\hat{H}_{\rm E}(\hat{r},\hat{p},\hat{R},\hat{P},t) \,,
\end{eqnarray}
where $\rm S$, $\rm B$ and $\rm C$ are subscripts denoting the subsystem, bath
and the coupling respectively. The Hamiltonian $H_{\rm E}(t)$ describes an external
time-dependent field which may interact with both the subsystem and bath.
The lower case coordinates describe
the subsystem degrees of freedom, while the upper case coordinates
describe the bath. 
The equation of motion for an arbitrary operator $\hat{\chi}(t)$ in the
Heisenberg picture is written in symplectic form as
\begin{eqnarray}
\frac{\partial}{\partial t}\hat{\chi}(t)=\frac{i}{\hbar}
\left[\begin{array}{cc}\hat{H}
& \hat{\chi}(t)\end{array}\right]
{\mathcal{B}}^c
\left[\begin{array}{c}\hat{H}\\
\hat{\chi}(t)\end{array}\right]\,,
\end{eqnarray}
where the matrix elements of the symplectic matrix~\cite{goldstein} are
defined as $B_{ij}^c=\epsilon_{ij}$, with $\epsilon_{ij}$ being the
complete antisymmetric tensor. 

It is assumed that the Hamiltonian of the bath depends on a pair of
canonically conjugate operators, $\hat{X} = (\hat{R}, \hat{P})$, and that the
coupling Hamiltonian $\hat{H}_{\rm C}$ depends only on the position coordinates
and not on the momenta.
The partial Wigner transform for the operator $\hat{\chi}$
is defined as
\begin{eqnarray}
\hat{\chi}_{\rm W}(X) = \int dz\, e^{i Pz/\hbar}\Big\langle R - \frac{z}{2}\Big|\hat{\chi}\Big|
R + \frac{z}{2}\Big\rangle\,,
\end{eqnarray}
where $X= (R,P)$ is the phase space point, defined in terms
of the canonically conjugate positions and momenta.

Upon taking the partial Wigner transform of the Heisenberg equation, 
one obtains
the Wigner-Heisenberg equation of motion:
\begin{eqnarray}
\frac{\partial}{\partial t}\hat{\chi}(X,t) = \frac{i}{\hbar}\left[\begin{array}{cc}
\hat{H}_{\rm W}(X) & \hat{\chi}_{\rm W}(X,t)\end{array}\right]{\mathcal{D}}
\left[\begin{array}{c}\hat{H}_{\rm W}(X) \\ \hat{\chi}_{\rm W}(X,t)\end{array}\right]\,,
\label{eq:WH-equation-of-motion}
\end{eqnarray}
where
\begin{eqnarray}
{\mathcal{D}} = \left[\begin{array}{cc} 0 & e^{\frac{i\hbar}{2}\stackrel{\leftarrow}
\partial_{k}{\mathcal{B}}^{c}_{kj}\stackrel{\rightarrow}\partial_{j}} \\ 
-e^{\frac{i\hbar}{2}\stackrel{\leftarrow}\partial_{k}{\mathcal{B}}^{c}_{kj}
\stackrel{\rightarrow}\partial_{j}} & 0 \end{array}\right]\,.
\end{eqnarray}
The symbols $\stackrel{\leftarrow}\partial_{k} = \stackrel{\leftarrow}\partial/\partial X_{k}$
and $\stackrel{\rightarrow}\partial_{k} = \stackrel{\rightarrow}\partial/\partial X_{k}$ denote
the operators of derivation with respect to the phase-space coordinates acting to the left and
right respectively.
Summation over repeated indices is implied.

The partial Wigner-transformed Hamiltonian takes the form:
\begin{eqnarray}
\hat{H}_{\rm W}(X,t)=\hat{H}_{\rm S}+H_{\rm B,W}(X)+\hat{H}_{\rm C,W}(R) 
+\hat{H}_{\rm E,W}(X,t)\,.
\label{eq:H_W-total}
\end{eqnarray}
When the Hamiltonians
$\hat{H}_{\rm B,W}$,
$\hat{H}_{\rm C,W}$ and $\hat{H}_{\rm E,W}(X,t)$ are at most quadratic
in $R$ and $P$, the action
of the terms in the matrix $\mathcal{D}$ are equivalent to their linear order
Taylor series expansion, since any higher order terms acting upon the
Hamiltonian will
yield zero. Under these conditions, algorithms for
simulating exact quantum dynamics in the partial Wigner representation
can be devised.

The above theory is also applicable in the case when the
external field Hamiltonian, $\hat{H}_{\rm E}$, depends
upon the bath coordinates, however, in the following we restrict
our study to situations where
it depends only upon the subsystem coordinates.


\section{Representation in a time-dependent basis}
\label{sec:sec3}

When considering Eq.~(\ref{eq:WH-equation-of-motion}), one can define
the following time-dependent Hamiltonian 
\begin{eqnarray}
\hat{h}_{\rm W}(R,t) 
&=&\hat{H}_{\rm S}+V_{\rm B,W}(R)+\hat{H}_{\rm C,W}(R)+\hat{H}_{\rm E,W}(t)\;,
\label{eq:h_W}
\end{eqnarray}
where $V_{\rm B,W}(R)$ is the potential energy of the bath.
A time-dependent basis can be defined in terms of the eigenstates
of $\hat{h}_{\rm W}(R,t)$:
$\hat{h}_{\rm W}(R,t)|\alpha; R,t\rangle = E_{\alpha}(R,t)|\alpha; R,t\rangle$.
In this basis the quantum evolution takes the form
\begin{eqnarray}
\chi_{\rm W}^{\alpha\alpha'}(X,t) =
{\cal T}\left\{
\sum_{\beta\beta'}\left(e^{i\int_{t_0}^t d\tau
\mathcal{L}^{t}(\tau)}\right)_{\alpha\alpha',\beta\beta'}\right\}
\chi_{\rm W}^{\beta\beta'}(X,t_0)\,,
\label{eq:quant-evolution}
\end{eqnarray}
where $t_0$ is the initial time, the symbol $\cal T$ denotes
time-ordering, and
\begin{eqnarray}
i{\mathcal{L}}_{\alpha\alpha'\beta\beta'}^{t}(t)
=i\tilde{\mathcal{L}}_{\alpha\alpha'\beta\beta'}(t)
+J_{\alpha\alpha'\beta\beta'}^{t}(t)\;.
\label{eq:calL^t}
\end{eqnarray}
In Eq.~(\ref{eq:calL^t}), we have introduced the transition operator
\begin{eqnarray}
J^{t}_{\aabb} = \langle\dot\alpha|\beta\rangle\delta_{\alpha'\beta'} 
+
\langle\beta|\dot\alpha'\rangle\delta_{\ab}\;,
\end{eqnarray}
which arises explicitly from the time dependence of the basis.
The operator $J_{\aabb}^t(t)$
determines the quantum transitions of the subsystem 
caused by the interaction with the external field.
The time-dependent Liouville operator $i\tilde{\mathcal{L}}_{\alpha\alpha'\beta\beta'}$
is similar in form to those first obtained in Refs.~\cite{sergi-theor-chem,mqc}
\begin{eqnarray}
i\tilde{\mathcal{L}}_{\alpha\alpha',\beta\beta'} 
&=& i\tilde{\mathcal{L}}^{0}_{\alpha\alpha'}
\delta_{\alpha\beta}\delta_{\alpha'\beta'}
+ \tilde{J}_{\alpha\alpha',\beta\beta'}(t)\,,
\end{eqnarray}
with $i\tilde{\mathcal{L}}_{\alpha\alpha'}^0=
i\tilde{\omega}_{\alpha\alpha'}(t)
+i\tilde{L}_{\alpha\alpha'}(t)$, and the
Bohr frequency given by
$\tilde\omega_{\alpha\alpha'}(R,t)=(E_{\alpha}(R,t)-E_{\alpha'}(R,t))/\hbar$.
The Liouville operator for the bath degrees of freedom
is given by
$i\tilde{L}_{\alpha\alpha'}=(P/M)\cdot(\partial/\partial R)
+(1/2)(\tilde{F}^{\alpha}_{\rm W}(t)+\tilde{F}^{\alpha'}_{\rm W})
\cdot\partial/\partial P$,
where $\tilde{F}^{\alpha}_{\rm W}(R,t)$ is a time-dependent 
Hellman-Feynman force for the
energy surface $E_{\alpha}(R,t)$.
The quantum transition operator is also similar in form
to that given in Ref.~\cite{mqc}:
\begin{eqnarray}
\tilde{J}_{\aabb}(t)=\tilde{\mathcal{T}}_{\alpha\rightarrow\beta}(t)
\delta_{\alpha'\beta'}
+\tilde{\mathcal{T}}_{\alpha'\rightarrow\beta'}^*(t)
\delta_{\alpha\beta}\,,
\end{eqnarray}
with
\begin{eqnarray}
\tilde{\mathcal{T}}_{\alpha\rightarrow\beta}(t)&=&
\frac{P}{M}\cdot d_{\alpha\beta}(R,t)\left(
1+\half\frac{\Delta E_{\alpha\beta}(t)d_{\alpha\beta}(R,t)}{\frac{P}{M}
\cdot d_{\alpha\beta}(R,t)}
\partialP\right)\,,
\\
\tilde{\mathcal{T}}^{*}_{\alpha'\rightarrow\beta'}
&=&\frac{P}{M}\cdot d^{*}_{\alpha'\beta'}(R,t)\left(
1+\half\frac{\Delta E_{\alpha'\beta'}(t)d_{\alpha'\beta'}^*(R,t)}
{\frac{P}{M}\cdot d^{*}_{\alpha\beta}(R,t)}
\partialP\right)\,,
\end{eqnarray}
and
$\Delta E_{\alpha\beta}(t) = E_{\alpha}(R,t) - E_{\beta}(R,t)$.
In the above, the coupling vector for the time-dependent states
has been introduced as
$d_{\alpha\beta}(R,t)=\langle\alpha;R,t|
\overrightarrow{\partial}/\partial R|\beta;R,t\rangle$.
 
The operator $\tilde{J}_{\aabb}(t)$ describes the quantum transitions
arising from the interaction between the system and the bath.
If such an interaction is weak,
the effect of $\tilde{J}_{\aabb}(t)$ is negligible.


\section{Metamolecule model}
\label{sec:sec4}

The metamolecule model considered in this work
comprises a two-level system
(the QD) coupled to a single resonating mode (RM), 
and subjected to a time-dependent external field. At this
point, a single harmonic mode was considered, as the computational
resources required by the phase-space grid algorithm, with which results
were being compared, rises exponentially with the dimension of the bath.
In the following, we will use adimensional coordinates and
parameters; they are expounded in detail in App.~\ref{app:app1}.

The QD Hamiltonian is
\begin{equation}
\hat{H}_{\rm S} = -\frac{\Omega}{2}\hat\sigma_{z}\;.
\label{eq:H_S}
\end{equation}
The resonant single-mode Hamiltonian, describing the MNP,
is defined as
\begin{equation}
H_{\rm B,W}=\frac{P^2}{2}+ \half \omega^2 R^2\;,
\label{eq:H_B}
\end{equation}
while the coupling to the RM is 
\begin{equation}
\hat{H}_{\rm C,W} = -c R\hat\sigma_{x}\;,
\label{eq:H_C}
\end{equation}
where $c$ is a coupling constant.
The external driving field is represented through the Hamiltonian
\begin{equation}
\hat{H}_{\rm E}(t)= g\cos(\omega_{d}t)\hat\sigma_{x}\,,
\label{eq:H_E}
\end{equation}
where $g$ denotes the driving strength of the external field,
and $\omega_{d}$
is the driving frequency. 
The symbols $\hat\sigma_x$ and $\hat\sigma_z$ denote the Pauli matrices.
The total Hamiltonian is given by Eq.~(\ref{eq:H_W-total}).

The energy eigenvalues of the Hamiltonian $\hat{h}_{\rm W}(R,t)$
in Eq.~(\ref{eq:h_W}) are
\begin{eqnarray}
E_{1,2}(R,t) &=& \Vb \pm \sqrt{\qomega + \gamma^2 + \gcossq + 2\gamma\gcos}\,,
\end{eqnarray}
where $\gamma = -cR$. In the basis of $\hat{h}_{\rm W}(R,t)$, 
the dynamics of arbitrary quantum operators
is defined by Eq.~(\ref{eq:quant-evolution}).
One can discretize time and obtain the evolution equation
\begin{eqnarray}
\chi_{\rm W}^{\alal}(t) = \sum_{\beta\beta'}{\mathcal{T}}
\bigg\{\exp\left[i\sum_{n}\tau_{n}
{\mathcal{L}}^{t}(\tau_{n})\right]\bigg\}_{\aabb}\chi_{\rm W}^{\beta\beta'}(t_{0})\,,
\end{eqnarray}
where $\sum_{n}\tau_{n} = t - t_0$. Using very small time steps $\tau_n$ and
the Dyson identity, one obtains
\begin{eqnarray}
\chi_{\rm W}^{\alal}(t)&=&\sum_{\beta\beta'}{\mathcal{T}}\prod_{n}
\Big\{\exp\left[i\tau_{n} \tilde{\mathcal{L}}^{0}_{\alal}(\tau_{n})\right]
\nonumber\\
&\times&
\left(1+\tau_{n}\tilde{J}_{\aabb} +\tau_{n}J^{t}_{\aabb}\right)\Big\}
\chi_{\rm W}^{\beta\beta'}(t_0)\,.
\end{eqnarray}
In the case of weak coupling to the bath, the action of
$\tilde{J}_{\aabb}$ can be disregarded. For $\tau_n = \tau$ for every $n$,
one then obtains:
\begin{eqnarray}
\chi_{\rm W}^{\alal}(R,P,t)
&=&
\sum_{\beta\beta'}{\mathcal{T}}\prod_{n}\Big\{\exp\left[i\tau
\tilde{\mathcal{L}}^{0}_{\alal}(\tau)\right]\left(1+\tau J^{t}_{\aabb}\right)\Big\}
\chi_{\rm W}^{\beta\beta'}(t_0)\;.
\label{eq:chi-discretised-weak-coupling}
\end{eqnarray}
Equation~(\ref{eq:chi-discretised-weak-coupling}) can be implemented by means of a 
stochastic algorithm. One can sample with probability $1/2$ one of the two terms in 
$J^{t}_{\aabb}$ acting at each time step. For each phase space point $(R,P)$,
one propagates a single deterministic step, dictated by $i\tilde{\cal L}_{\aabb}^0(t)$.
At the end of such a step, the quantum transition, due to
the external field, is sampled.
Transition probabilities can be defined as:
\begin{eqnarray}
{\mathcal{P}}_{\beta\rightarrow\alpha} = \frac{\tau| \langle\dot{\alpha}|\beta\rangle |}
{1 + \tau|\langle\dot{\alpha}|\beta\rangle|}\,.
\label{eq:transition}
\end{eqnarray}
The probability of rejecting the transition will then be given by
\begin{eqnarray}
{\mathcal{Q}}_{\beta\rightarrow\alpha} = \frac{1}{1+\tau|\langle\dot{\alpha}|\beta\rangle|}\,.
\label{eq:no-transition}
\end{eqnarray}
For the two-level model that we are studying, the eigenstates can be calculated exactly
and therefore, so can the transition probabilities in Eqs.~(\ref{eq:transition})
and~(\ref{eq:no-transition}).


\section{Results}
\label{sec:sec5}

The initial state of the system is defined as
\begin{eqnarray}
\hat{\rho}_{\rm W}(R,P) = \left(\begin{array}{cc} 0\,\,\,\, & 0 \\ 0\,\,\,\, & 1\end{array}\right)\times \rho_{\rm B,W}(R,P)\,,
\label{eq:inirho}
\end{eqnarray}
where the $\rho_{\rm B,W}(R,P)$ is the Wigner function for the bath, given by
\begin{eqnarray}
\rho_{\rm B,W}(R,P) = \frac{\tanh(\beta\omega/2)}{\pi}\exp\left[-\frac{2\tanh(\beta\omega/2)}
{\omega}\left(\frac{P^{2}}{2}+\frac{\omega^{2}R^{2}}{2}\right)\right]\,,
\end{eqnarray}
and the matrix on the right hand side of Eq.~(\ref{eq:inirho}) is given
in the subsystem basis.
The average values of an arbitrary phase-space dependent operator,
$\hat{\chi}_{\rm W}(R,P,t)$, are calculated as
\begin{equation}
\langle\hat{\chi}_{\rm W}(R,P,t)\rangle
={\rm Tr}'\int dRdP \hat{\rho}_{\rm W}(R,P) \hat{\chi}_{\rm W}(R,P,t) \;.
\label{eq:averages}
\end{equation}
The partial trace and the evolution of $\hat{\chi}_{\rm W}(R,P,t)$
are calculated in the time-dependent basis and with the
algorithm sketched in Secs.~\ref{sec:sec3} and~\ref{sec:sec4}.

In order to determine the effect of the external field upon the metamolecule,
we kept the values of the system parameters unchanged 
while varying the coupling strength of the external field.
The values of the system parameters are $\beta = 12.5$, $c = 0.01$,
$\Omega = 0.8$, $\omega = 0.5$, and $\omega_{d} = 0.05$.
These values lead to a weak coupling between
the QD and the RM, so that the quantum transitions caused by the interaction with the resonant mode
can be neglected. The values for the
coupling strength of the external field used in this study were
$g = 0.1, 0.3, 0.5, 0.7, 0.9, 1.5$.
However, 
in order to demonstrate the differences between weak and strong coupling,
only the results for $g = 0.1$ and $g = 1.5$ are shown in the figures.
In the case of the piece-wise deterministic algorithm, described in Secs.~\ref{sec:sec3}
and~\ref{sec:sec4},
a time step of $\tau = 0.1$ was employed, and a total of $10^5$ trajectories were
propagated in each calculation.
Instead, the phase-space grid algorithm (described in App.~\ref{app:app1})
requires a smaller time step of $\tau = 0.001$.
The phase-space grid spacing was $\Delta R = \Delta P = 0.1$.

\begin{figure}[h]
\resizebox{10.0cm}{6.0cm}{
\includegraphics{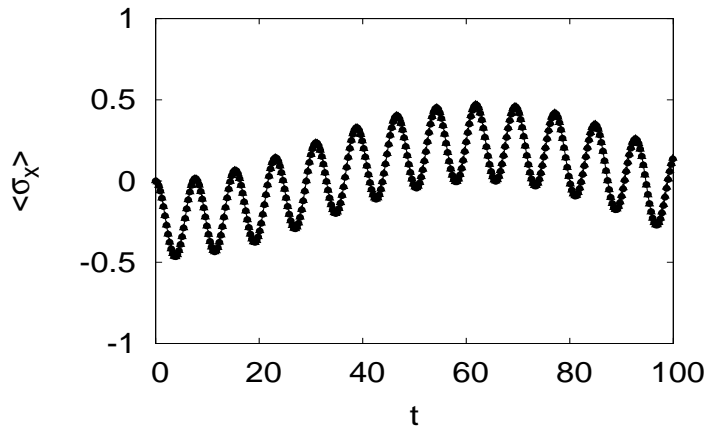}
}
\caption{
Plot of the average coherence, $\langle\sigma_{x}\rangle$,
as a function of time.
The solid black circles refer to the results obtained with
the piece-wise deterministic algorithm while
the white triangles refer to those of the grid integration.
A continuous line connects the data points of the piece-wise deterministic
algorithm. The values of the system parameters are
$\beta=12.5$, $c=0.01$, $\Omega = 0.8$, $\omega = 0.5$, 
$\omega_d = 0.05$ and $g = 0.1$,
corresponding to weak driving field strength. 
The results of the two algorithms are indistinguishable to the human eye.
}
\label{fig:fig1}
\end{figure}

In Fig.~\ref{fig:fig1}, a comparison of the results obtained with the two
different algorithms, for the average coherence 
of the quantum dot as a function of time, is shown.
With a value of $g = 0.1$, this calculation corresponds to a weak driving field.
The two results agree almost exactly, so that they cannot be
distinguished by the human eye.
Moreover, the error bars are negligible.
In this weak coupling case, the oscillations remain relatively small, with 
the values of $\langle\sigma_{x}\rangle$ ranging between $-0.5$ and $0.5$. 
A very slow mode of oscillation, with angular frequency $\approx 0.05$
appears to be superimposed to a fast mode with angular frequency $\approx 0.81$.
The slow frequency is basically that of the driving field
while the slow one corresponds to the tunnel splitting, shifted by
a very small amount because of the weak coupling to the field.
As expected, in the case of weak coupling to the external field
it is the tunnel splitting that dominates the evolution in time
of $\langle\sigma_x\rangle$.

\begin{figure}[h]
\resizebox{10.0cm}{6.0cm}{
\includegraphics{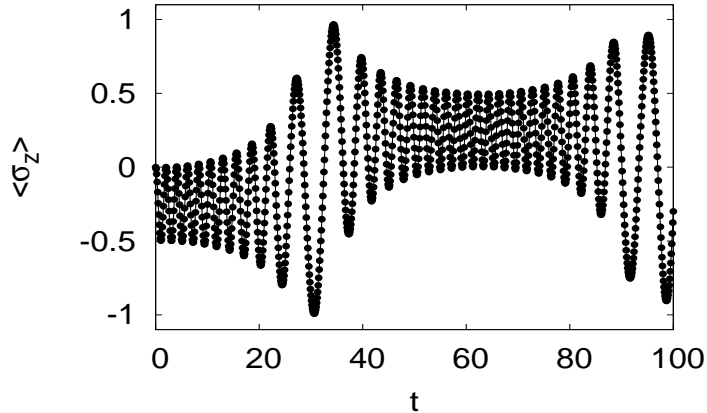}
}
\caption{
Plot of the average coherence, $\langle\sigma_{x}\rangle$,
as a function of time.
The solid black circles refer to the results obtained with
the piece-wise deterministic algorithm while 
the white triangles refer to those of the grid integration.
A continuous line connects the data points of the piece-wise deterministic 
algorithm. 
The values of the system parameters are $\beta=12.5$, $c=0.01$, $\Omega = 0.8$, $\omega = 0.5$,
$\omega_d = 0.05$ and $g = 1.5$, corresponding to
strong driving field strength.
The results of the two algorithms are indistinguishable to the human eye.
}
\label{fig:fig2}
\end{figure}

Figure~\ref{fig:fig2} shows the results of the calculations with $g = 1.5$,
corresponding to a strong driving field.
The results produced by the two algorithms are indistinguishable also in this
case, with error bars remaining smaller than the points
for the entire simulation time.
In this case, the evolution of the coherence in time displays
a non-linear pattern: it starts with fast oscillations around
$\langle\sigma_x\rangle=-0.25$ from $t=0$ to $t\approx20$;
within $20<t<40$ it undergoes large oscillations, and then it switches
to oscillating fast around the value of $\langle\sigma_x\rangle=0.25$;
it does so until $t=80$, when the large oscillations start again.
We can therefore conclude that the QD switches between
two different dynamical regimes, one where the coherence is positive,
and the other where it is negative. 
We can associate a period to such a switching dynamics,
whose numerical value matches that of the driving field.
Because of the strong coupling, the frequency of the fast oscillations is
about 200\% greater than that obtained in the case of weak coupling to
the external field. The inspection of Fig.~\ref{fig:fig2} reveals
that the dynamics of $\langle\sigma_x\rangle$ is now dominated
by the strong coupling to the driving field.

\begin{figure}[h]
\resizebox{10.0cm}{6.0cm}{
\includegraphics{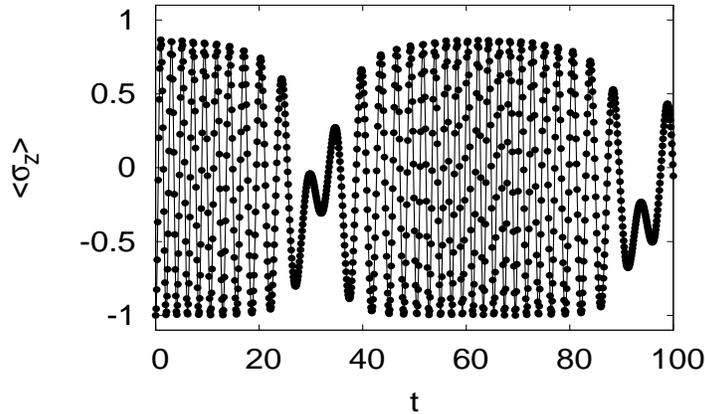}
}
\caption{
Plot of the average population difference, $\langle\sigma_{z}\rangle$,
as a function of time.
The solid black circles refer to the results obtained with
the piece-wise deterministic algorithm while
the white triangles refer to those of the grid integration.
A continuous line connects the data points of the piece-wise deterministic
algorithm.
The values of the system parameters are $\beta=12.5$, $c=0.01$, $\Omega = 0.8$, $\omega = 0.5$,
$\omega_d = 0.05$ and $g = 1.5$, corresponding to
strong driving field strength.
The results of the two algorithms are indistinguishable to the human eye.
}
\label{fig:fig3} 
\end{figure}

In Fig.~\ref{fig:fig3}, we show the evolution in time
of the average of the population difference in the strong driving regime. 
Such a quantity also shows a non-linear pattern,
with regions of large and fast oscillations separated by regions of slow and
small oscillations.
At weak driving strengths, this behavior is not noticeable,
and the oscillations are much smaller.
This corresponds to a lower percentage of the ensemble
of trajectories of the piece-wise deterministic algorithm
being driven into the excited state by the external field.

\begin{figure}[h]
\resizebox{10.0cm}{6.0cm}{
\includegraphics{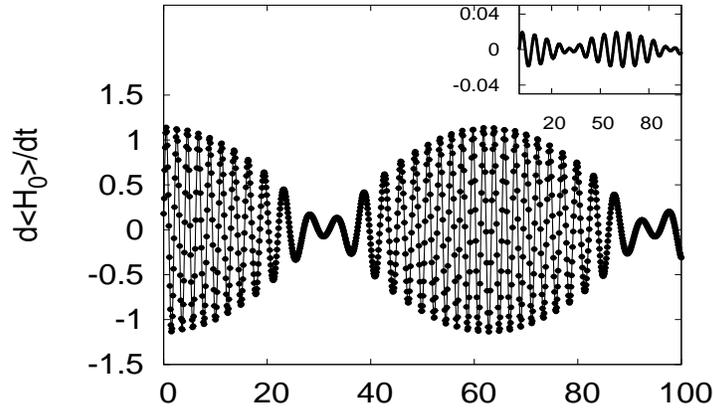}
}
\caption{
Plot of the rate of change of the expectation value of the Hamiltonian
for the metamolecule. The main figure displays the results
for the strong driving field strength while the inset shows
those for the weak driving field strength.
The continuous line connects the data points.
}
\label{fig:fig4}
\end{figure}

In Fig.~\ref{fig:fig4}, we plot the rate of change of the expectation value
of the energy of the QD-RM metamolecule in two different cases:
$g = 0.1$ and $g = 1.5$.
For $g=1.5$, such a quantity shows a non-linear pattern,
with regions of large and fast oscillations separated by regions of slow and
small oscillations, in the same time intervals where $\langle\sigma_x\rangle$
and $\langle\sigma_z\rangle$ do.
Instead, for $g = 0.1$  the rate of change of the average energy of 
the metamolecule
has significantly smaller oscillations around the mean
and the structured pattern is almost absent.
The calculation of the rate of change of the energy of RM was also performed,
and it was found that the RM reacts slowly to the change in energy of the QD.
This justifies
the neglect of the quantum transitions due to the coupling to the RM.


\section{Conclusions}
\label{sec:sec6}

Employing the partial Wigner representation of quantum mechanics,
we have studied the dynamics of a model for a quantum dot
coupled to a single resonating mode of a metal nano-particle.
We have treated the case in which the
Hamiltonian is explicitly time-dependent, due to the presence of
a driving field directly coupled to the quantum dot.
An explicitly time-dependent basis has been used for the
representation of the equations of motion and generalized propagation
schemes have been devised both in terms of a piece-wise deterministic
algorithm and of a grid-based numerical integration.
The results obtained by using these two algorithms
were compared.
We have shown that both schemes of integration produce numerically
indistinguishable results.
However, the piece-wise deterministic algorithm has the definite 
advantage of being able to treat systems with
a higher number of discrete energy levels (such as three- or
four-level quantum dots) and many more (hundreds or thousands)
of resonating modes in an affordable computational time (see, for example,
Ref.~\cite{num-app-quant}).

We have studied the effect of the driving strength  of the external field
upon the quantum dot.
By monitoring the average of the time variation of the energy
of the metamolecule model, as well as the coherence and 
the population difference of the quantum dot,
we observe distinct non-linear behavior in the case of strong coupling
to the driving field.

Both the algorithms presented in this work and the results obtained
can be considered as a first step toward the development
of an effective approach (alternative to the use of
master equations) for studying plasmonic metamolecules.


\appendix

\section{Adimensional coordinates and parameters}
\label{app:app1}

Upon indicating the dimensional coordinates and parameters
with a prime, and introducing the energy scale
$\hbar\omega_{\rm a}'$, we have the following definitions:
\begin{eqnarray}
\Omega&=&\frac{\Omega'}{\omega_{\rm a}'} \;, \\
P&=&\frac{P'}{\sqrt{M'\hbar\omega_{\rm a}'} } \;, \\
R&=&\sqrt{\frac{\hbar \omega_{\rm a}'}{\hbar}} R' \;, \\
\omega&=&\frac{\omega'}{\omega_{\rm a}'}  \;,\\
c&=& \frac{c'}{\sqrt{ \hbar\omega_{\rm a}^{\prime 3}M'}}  \;,\\
g&=&\frac{g'}{\hbar\omega_{\rm a}'}  \;,\\
\beta&=&\hbar\omega_{\rm a}'\beta'\;.
\end{eqnarray}
The symbol $M$ is the inertial parameter of the oscillator,
which has been set to unity.
Upon choosing $\omega_{\rm a}'$ in such a way that
the the frequency $\omega=0.5$ of the oscillator
in Eq.~(\ref{eq:H_B}) corresponds to the value
$\omega'=8.9 \times 10^{12}$ Hz, which is typical
for the dynamics of metal nano-particles~\cite{plasmonics-tame},
we obtain a spanned time-scale in our simulations of
$5.62 \times 10^{-12}$ s
and an energy variation for the quantum dot, as shown in Fig.~\ref{fig:fig4},
of $24.6$ meV.

\section{Phase-space Grid Algorithm}
\label{app:app2}

The equation of the density matrix in the partial Wigner representation
is analogous to that Given in Eq.~(\ref{eq:WH-equation-of-motion}):
\begin{eqnarray}
\frac{\partial}{\partial t}\hat{\rho}(X,t)=-\frac{i}{\hbar}\left[\begin{array}{cc}
\hat{H}_{\rm W}(X) & \hat{\rho}_{\rm W}(X,t)\end{array}\right]{\mathcal{D}}
\left[\begin{array}{c}\hat{H}_{\rm W}(X)
\\ \hat{\rho}_{\rm W}(X,t)\end{array}\right]\;.
\label{eq:WH-rho-equation-of-motion}
\end{eqnarray}
Using the basis defined by $\hat{H}_{\rm S}|\alpha\rangle = \epsilon_{\alpha}
|\alpha\rangle$ ($\alpha=1,2$),
Eq.~(\ref{eq:WH-rho-equation-of-motion}) can
be written as
\begin{eqnarray}
\partialt\rho_{\rm W}^{\alal}(R,P,t) &=& -i\tomega_{\alal}\rhowalal -L\rhowalal 
- \iohbar\left(H_{\rm C,W}^{\alpha\beta}\rho_{\rm W}^{\beta\alpha'} 
- \rho_{\rm W}^{\alpha\beta'}H_{\rm C,W}^{\beta'\alpha'}\right) 
\nonumber
\\
&-& \iohbar\left(H_{E}^{\alpha\beta}\rho_{\rm W}^{\beta\alpha'} 
- \rho_{\rm W}^{\alpha\beta'}H_{E}^{\beta'\alpha'}\right)
\nonumber
\\
&+& \half\left(\frac{\partial H_{\rm C,W}^{\alpha\beta}}{\partial R}\frac{\partial\rho_{\rm W}^{\beta\alpha'}}
{\partial P} + \frac{\partial\rho_{\rm W}^{\alpha\beta'}}{\partial P}\frac{\partial H_{\rm C,W}^{\beta'\alpha'}}
{\partial R}\right)\;,
\label{eq:qcle-eq-sub}
\end{eqnarray}
where the frequency $\tomega_{\alal} = \left(\epsilon_{\alpha} - \epsilon_{\alpha'}\right)/\hbar$
has been defined. The Liouville operator, $L$, is given by
\begin{eqnarray}
L = P\partialR - \frac{\partial H_{B,W}}{\partial R}\partialP\;.
\end{eqnarray} 
One can introduce the following frequencies $\tilde{\omega}^{\alal}$:
\begin{eqnarray}
\label{eq:frequencies}
&&\tilde{\omega}^{11} = 0, \hspace{1.3cm} \tilde{\omega}^{22} = 0,
\nonumber
\\
&&\tilde\omega^{12} = \Omega, \hspace{1cm} \tilde\omega^{21} = -\hbar\Omega.
\end{eqnarray}
The equations
of motion for matrix elements of the density operator can be written
explicitly as:
\begin{eqnarray}
\partialt\rhooo(R,P,t) &=& 
\iohbar cR\left(2i\text{Im}\left[\rhoto\right]\right) -\iohbar
\gcoswt\left(2i\text{Im}\left[\rhoto\right]\right)
\nonumber
\\
&-& L\rhooo - 
\frac{c}{2}\partialP\left(2\text{Re}\left[\rhoto\right]\right)\;,
\\
\partialt\rhoto(R,P,t) &=&
-i\tomega_{21}\rhoto + \iohbar cR\left(\rho_{\rm W}^{11} - \rho_{\rm W}^{22}\right)
-\iohbar\gcoswt\left(\rhooo - \rhott\right) 
\nonumber
\\
&-& L\rhoto
- \frac{c}{2}\partialP\left(\rhooo +\rhott\right)\;,
\\
\partialt\rhott(R,P,t) &=&
-\iohbar cR\left(2i\text{Im}\left[\rhoto\right]\right) 
+\iohbar\gcoswt\left(2i\text{Im}\left[\rhoto\right]\right)
\nonumber
\\
&-& L\rhott -
\frac{c}{2}\partialP\left(2\text{Re}\left[\rhoto\right]\right)\,.
\end{eqnarray}
In order to simplify the integration, one can use the
definition
\begin{equation}
\rho_{\rm W}^{\alal}(X,t)=\eta_{\rm W}^{\alal}(X,t)
e^{-i\tilde\omega^{\alal}t}\;.
\end{equation}
Hence, using dimensionless coordinates,
the equations of motion become
\begin{eqnarray}
\partialt\etaoo &=& -2cR\left(-\text{Re}\left[\etato\right]\sin\left(\tomega_{21}t\right) +
\text{Im}\left[\etato\right]\cos\left(\tomega_{21}t\right)\right) 
\nonumber
\\
&+& 2g\coswt\left(-\text{Re}\left[\etato\right]\sin\left(\tomega_{21}t\right) +
\text{Im}\left[\etato\right]\cos\left(\tomega_{21}t\right)\right)
\nonumber
\\
&-& L\etaoo
- c\partialP\left(\text{Re}\left[\etato\right]\cos\left(\tomega_{21}t\right) + 
\text{Im}\left[\etato\right]\sin\left(\tomega_{21}t\right)\right)\;,
\label{eq:pde-11}
\\
\partialt\etatt &=& 2cR\left(-\text{Re}\left[\etato\right]\sin\left(\tomega_{21}t\right) +
\text{Im}\left[\etato\right]\cos\left(\tomega_{21}t\right)\right)
\nonumber
\\
&-& 2g\coswt\left(-\text{Re}\left[\etato\right]\sin\left(\tomega_{21}t\right) +
\text{Im}\left[\etato\right]\cos\left(\tomega_{21}t\right)\right)
\nonumber
\\
&-& L\etatt
- c\partialP\left(\text{Re}\left[\etato\right]\cos\left(\tomega_{21}t\right) + 
\text{Im}\left[\etato\right]\sin\left(\tomega_{21}t\right)\right)\;,
\label{eq:pde-22}
\\
\partialt\left(\text{Re}\left[\etato\right]\right) &=& 
-cR\left(\etaoo - \etatt\right)\sin(\tomega_{21}t)+\gcoswt\left(\etaoo - \etatt\right)\sin(\tomega_{21}t)
\nonumber
\\
&-& L\left(\text{Re}\left[\etato\right]\right)
- \frac{c}{2}\partialP\left(\etaoo + \etatt\right)\cos(\tomega_{21}t)\;,
\label{eq:pde-21re}
\\
\partialt\left(\text{Im}\left[\etato\right]\right) &=&
cR\left(\etaoo - \etatt\right)\cos(\tomega_{21}t) -\gcoswt\left(\etaoo - \etatt\right)\cos(\tomega_{21}t)
\nonumber
\\
&-& L\left(\text{Im}\left[\etato\right]\right)
- \frac{c}{2}\partialP\left(\etaoo + \etatt\right)\sin(\tomega_{21}t)\;.
\label{eq:pde-21im}
\end{eqnarray}
Equations~(\ref{eq:pde-11}), (\ref{eq:pde-22}), (\ref{eq:pde-21re}) and (\ref{eq:pde-21im})
describe a set of coupled partial differential equations (PDE's)
which can be solved to obtain the elements of the density matrix as functions of time. In order
to solve these coupled PDE's, the numerical integration approach known as the method of lines
can be employed. The method of lines transforms the PDE's into a set of ordinary differential
equations (ODE's) by using finite difference approximations for all but one of the integration variables.
In this case, the phase-space derivatives are approximated, while the time variable is left
un-approximated. For the types of potentials studied in this work, a fourth-order finite difference
approximation for the phase-space derivatives proves to be more than sufficient:

\begin{eqnarray}
\frac{\textrm{d}f}{\textrm{d} X} = \left(\frac{f(X - 2\derivd X) - 8f(X - \derivd X) 
+ 8f(X + \derivd X) - f(X + 2\derivd X)}{12\derivd X}\right)\,.
\end{eqnarray}
Once the conversion from PDE's to ODE's has been performed, a numerical integration method
can be utilized. In this work, a Runge-Kutta 5 Cash-Karp method was found to be suitable for
stable numerical results.
In such an approach, the phase-space of the system is essentially discretized into a numerical
grid, upon which the coupled ODE's are solved.

\begin{acknowledgements}
We are grateful to Prof. Mark Tame for useful discussions and
for having stimulated our interest in quantum plasmonics.

This work is based upon research supported by
the National Research Foundation of South Africa.
\end{acknowledgements}

\end{document}